%% file: Arxiv.tex
\documentclass[12pt]{article}

\usepackage{spaccent}
\usepackage{fullpage,multicol}
\usepackage{hyperref}
\usepackage{graphicx}
\usepackage{amssymb}
\usepackage{amsmath}
\usepackage{latexsym}
\usepackage{color}
\usepackage[english]{babel}
\usepackage{comment}
\usepackage{dirtytalk}
\usepackage{dblfloatfix}
\usepackage{ntheorem}
\usepackage{siunitx}
\newcommand{\sixs}[0]{\unit[unit-font-command = \mathit]{\second}}

\newcommand{\sixhz}[0]{\unit[unit-font-command = \mathit]{\hertz}}
\newcommand{\sixbits}[0]{\unit[unit-font-command = \mathit]{bits}}

\newcommand{\sixgigabytes}[0]{\unit[unit-font-command = \mathit]{Gb}}

\theoremseparator{:}

\usepackage{tocloft}
\usepackage{enumitem}
\usepackage[dvipsnames]{xcolor}
\usepackage{pifont}
\usepackage{multicol}

\newcommand{\ignore}[1]{}

\newcommand{\blue}[1]{\textcolor{blue}{#1}}

\renewcommand{\blue}[1]{\textcolor{black}{#1}}

\usepackage{lipsum}
\newcommand\blfootnote[2]{%
  \begingroup
  \renewcommand\thefootnote{#2}\footnote{#1}%
  \addtocounter{footnote}{-1}%
  \endgroup
}
\newcommand{\yesj}[0]{\textcolor{OliveGreen}{\footnotesize\ding{52}}}
\newcommand{\noj}[0]{}

\begin{document}
\specialaccent
\thispagestyle{empty}



\begin{center}
\vspace{4mm}
\textbf{\large \blue{Effectively} obtaining acoustic, visual and textual data from videos}

\vspace{2mm}
\textbf{Jorge E. León}\textsuperscript{1}\blfootnote{E-mail: jorgleon@alumnos.uai.cl}{*} and \textbf{Miguel Carrasco}\textsuperscript{2}

\vspace{2mm}
\footnotesize{\textsuperscript{1} Adolfo Ibánez University (UAI), Santiago, Chile}

\footnotesize{\textsuperscript{2} Diego Portales University (UDP), Santiago, Chile}
\end{center}

\begin{abstract}
The increasing use of machine learning models has amplified the demand for high-quality, large-scale multimodal datasets.
However, the availability of such datasets, especially those combining \blue{acoustic}, visual and textual data, remains limited.
This paper addresses this gap by proposing a method to extract related audio-image-text observations from videos.
We detail the process of selecting suitable videos, extracting relevant data pairs, and generating descriptive texts using image-to-text models.
Our approach ensures a robust semantic connection between modalities, enhancing the utility of the created datasets for various applications.
We also discuss the challenges encountered and propose solutions to improve data quality.
The resulting datasets, publicly available, aim to support and advance research in multimodal data analysis and machine learning.

\textbf{Keywords:} Data generation, Multimodal data, Image, Audio, Text, Video.
\end{abstract}


\input{Publicacion/Introduccion}
\input{Publicacion/Related_work}
\input{Publicacion/Estado_del_Arte}
\input{Publicacion/Metodo}
\input{Publicacion/Dataset}
\input{Publicacion/Conclusiones}

\newpage
\bibliographystyle{plain}
\bibliography{citas.bib}


\end{document}

%% file: Publicacion/Introduccion.tex
\section{Introduction}\label{sec:Introducción}

\begin{figure*}[t]
\centering
\includegraphics[width=16.4cm]{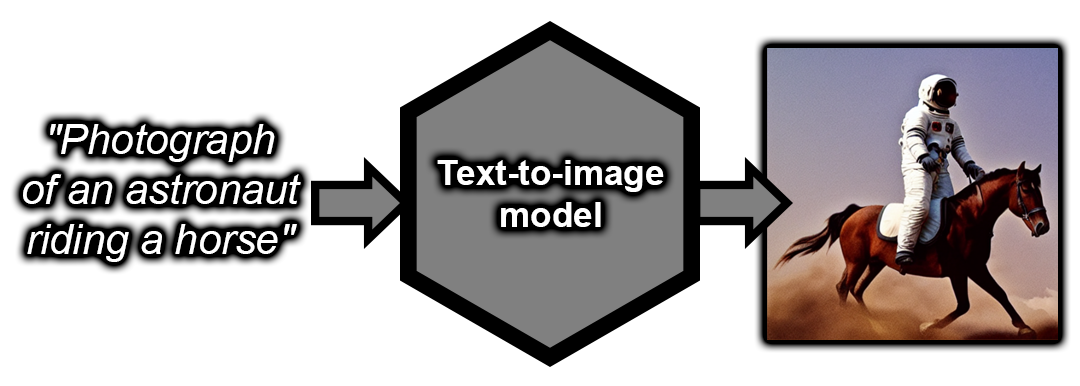}
\caption{Text-to-image generation example. Text-to-image is a technique that generates images from textual descriptions, allowing users to create visual content based on their written prompts. Some popular models that perform this task are Stable Diffusion \cite{High-Resolution_Image_Synthesis, SDXL, Stable_diffusion_3, Stable_Diffusion}, DALL·E \cite{Zero-Shot_Text-to-Image, Improving_Image_Generation_with_Better_Captions, DALLE_3}, Imagen \cite{Photorealistic_Text-to-Image_Diffusion_Models, Imagen3} and FLUX \cite{Flux}.}
\label{fig:texto-a-imagen}
\end{figure*}

In recent years, there has been an unprecedented development in the world of machine learning \cite{From_Classical_Machine_Learning_to_Deep_Neural_Networks}. Several models have begun to excel in creative activities (previously considered exclusive to human minds by many) \cite{Text-to-image_Diffusion_Models, Creativity_and_Machine_Learning}, and even using non-specialized hardware \cite{A_Survey_of_On-Device_Machine_Learning}.
In this scenario, models have emerged that can generate text associated with an image \cite{CLIP, BLIP, BLIP-2}; just as others have appeared that, based on texts/prompts, are capable of generating images that can fairly faithfully represent said texts \blue{\cite{Text-to-image_Diffusion_Models, SDXL, Imagen3, Flux}}. An example of this can be seen in Figure \ref{fig:texto-a-imagen}.

From this last task, usually referred to as text-to-image, several others emerge, such as: inpainting \cite{Image_inpainting}, outpainting \cite{Outpainting_Images_and_Videos}, or image-to-image \cite{Image-to-Image_Translation, Comparison_and_Analysis_of_Image-to-Image}.
Commonly, the conditioning of all the aforementioned tasks tends to be text-based, and there are a few popular datasets to train such models \cite{Microsoft_COCO, Laion-5b, Conceptual_12M, DeepFashion}. This is a simple example of multimodal data being used nowadays.

However, it is not unheard of to find undesirable entries, in any third-party dataset \cite{Large_image_datasets}, or a lack of datasets for specific tasks (e.g. medical image analysis \cite{A_survey_on_Image_Data_Augmentation_for_Deep_Learning}).
Similar inconveniences can also be found when dealing with datasets that include audio \cite{Audio_Language_Datasets}.
In particular, it has been mentioned that, relative to image datasets, audio-visual datasets are few and far between \cite{Audio-to-Image_Cross-Modal}.
Currently, this in turn can be explained by the apparent low motivation on exploring fields such as audio conditioned image-to-image \cite{Codi2, Any-to-Any_Generation, BindDiffusion}, in contrast with text conditioned image-to-image \blue{\cite{Image-to-Image_Translation, Text-to-image_Diffusion_Models, High-Resolution_Image_Synthesis, SDXL}}.
While there are numerous image-to-image works that condition the input image using text, there are not many that do so with audio (whether with or without added text involved) nor there are common guidelines to help researchers form these datasets on their own.
\blue{Additionally, as we will explain below, the option of adapting textual/visual models to work with acoustic inputs has significant drawbacks that discourage it, instead of directly training an acoustic model for the given task.}

\blue{It goes without saying that efforts in this topic} could have an impact on: multimodal data analysis, correction of low-quality/low-resolution recordings, video generation for various purposes (virtual assistants, music videos, video transitions, etc.), democratization of artificial intelligence, augmented reality that incorporates the user's environmental audio, transfer learning with multimodal models, among others \blue{\cite{Remote_Sensing_Image_Generation_From_Audio, Audio_deepfakes, Deep_Audio-visual_Learning, Deep_Audio-visual_Learning, A_Survey_on_Audio_Synthesis}}.

In light of the above and wanting to work with a specific type of \blue{acoustic}-visual data, we formalized a method to generate audio-image-text observations based on videos (including the textual modality, in order to expand the utility of our datasets), and employed it to generate the data we desired for our future research. This paper delves into all of that.

In summary, in this paper we address the need for high-quality, large-scale multimodal datasets that combine \blue{acoustic}, visual, and textual data (which are currently limited).
Keeping in mind the importance of maintaining a strict semantic connection between audio and visual data to improve dataset quality, as well as the ideal of minimizing data modality conversions to preserve data integrity and quality, we propose a coherent and systematic approach to extract audio-image-text observations from videos.
We discuss about our results, generating more than 2,000,000 audio-image pairs from over 280,000 videos, together with the transformation we utilized to obtain the respective texts and some pending challenges we encountered along the process.

%% file: Publicacion/Related_work.tex
\blue{\section{Related Work}\label{sec:Related_work}}

\begin{table}[t]
    \begin{tabular}{m{0.17\textwidth}m{0.21\textwidth}m{0.49\textwidth}m{0.0\textwidth}}
        \hline
        \centering \textbf{Task} & \centering \textbf{Description} & \centering \textbf{Nuances} &\\ \hline
        Image-to-audio & Based on an image, an audio is generated that conveys the same semantic information as the input image. & Advances have been made in the generation of audios that mimic the possible soundscape for a given image \cite{I_Hear_Your_True_Colors, Any-to-Any_Generation}. In a similar fashion, audios can also be generated from videos, which are nothing more than an ordered collection of images \cite{A_Survey_on_Audio_Synthesis, Deep_Audio-visual_Learning}. &\\ \hline
        Text-to-audio & Based on a text, an audio is generated that conveys the same semantic information as the input text. & Some models are able to resemble a human voice reading the text given as input (subtask usually referred to as text-to-speech \cite{Audio_deepfakes, A_Survey_on_Audio_Synthesis, Vall_e, Towards_audio_language_modeling}). Moreover, some even make music \cite{Mustango} and generate the lyrics based on text input \cite{Jukebox}, or generate sounds that accommodate to a given description \cite{Fugatto, AudioGen, Any-to-Any_Generation, AudioLDM}. &\\ \hline
        Audio-to-image & Based on an audio, an image is generated that conveys the same semantic information as the input audio. & Voice recordings can be used to condition the modification of human faces so their mouths adapt to the corresponding sounds (i.e. lip sync \cite{Audio_deepfakes, Multimodal_Image_Synthesis_and_Editing}), and even the whole face can be created from scratch with the aforementioned recordings \cite{A_Survey_on_Audio_Synthesis}. In addition, some models are capable of representing scenarios where a specific audio is produced \cite{Any-to-Any_Generation, Deep_Audio-visual_Learning}. &\\ \hline
        Audio-to-text & Based on an audio, a text is generated that conveys the same semantic information as the input audio. & The most popular subtask here probably is speech transcription (or recognition) \cite{Audio_deepfakes, Multimodal_Learning_With_Transformers, transcripter_generation, whisper}. However, models that remarkably generate text description (or captions) from audios in general have begun to arise in recent years \cite{Any-to-Any_Generation, AudioSetCaps, BLAP, BLAT}. &\\ \hline
    \end{tabular}
    \caption{A summary on the most common generative audio-text and audio-image tasks.}
    \label{tab:resumen_investigaciones}
\end{table}

Our literature review provided clear evidence on the existence of relationships between audio and text that represent the same situation, as well as between audio and image, that should be further exploited by research and modern models (for a small summary on generative tasks that involve said modality combinations, consult Table \ref{tab:resumen_investigaciones}).

\blue{Exploring the most relevant cases to image-to-image conditioned by audio,} there are some examples of image generation based on audio and text \cite{AudioToken, TimbreCLIP}, and there are even cases of image-to-image generation assisted only by audio, but for specific cases such as face changes (which replace a person's features with another's while maintaining consistency with the original voice recording) or lip synchronizations (where, for an image of a person, a video is generated while simulating mouth movement according to a voice recording) \cite{Audio_deepfakes, A_Survey_on_Audio_Synthesis}, which could be labeled more as a case of inpainting than image-to-image.
Finally, advances in other similar areas can also be highlighted (such as text-to-video, appreciable with models like Sora \cite{sora_analisis, sora_reporte}, Veo \cite{Veo}, Gen-3 \cite{Gen3} and Movie Gen \cite{Movie_Gen}), and more information on some of these developments can be found at \cite{A_Survey_on_Generative_Diffusion_Models, A_survey_of_multimodal_deep_generative_models}.

Currently, image-to-image generation conditioned by audio is a little explored area of high interest in the community. To the best of our knowledge, one of the best models to date for this task is the recent CoDi model \cite{Any-to-Any_Generation}.
This is a model that can take any combination of audio, image, text, and video inputs, and create material of any of those types (a task they called any-to-any).
Additionally, a new version (CoDi-2) has also been published, which is more flexible and adapted to conversations \cite{Codi2}.
Another similar option is NExT-GPT, which also allows for a conversational creative process, and it works as well with audio, image, text, and video inputs \cite{NExT_GPT}.
Despite their promising results for future iterations, they have not yet reached a quality that could be considered ideal.
Probably, the best open-source model for this task is BindDiffusion \cite{BindDiffusion}.
This model is both based on the image generation model Stable Diffusion \cite{High-Resolution_Image_Synthesis}, and on the multimodal encoder ImageBind, which incorporates six modalities, including, predictably, audio and image \cite{Imagebind}. 
Notwithstanding its apparently higher quality than CoDi or NExT-GPT, it also has room for improvement, and it is not evident that it is always advisable to include the largest possible number of data modalities in these models (as seems to have been attempted in all of these cases).

\blue{The datasets involved in the training of the three previous models also shed some light on the lack of and demand for more multimodal datasets.
For instance, CoDi needs to leverage different datasets (namely, LAION-400M \cite{LAION-400M}, AudioSet \cite{AudioSet}, AudioCaps \cite{AudioCaps}, Freesound 500K, BBC Sound Effect, SoundNet \cite{SoundNet}, WebVid-10M \cite{Frozen_in_Time}, and HD-VILA-100M \cite{Advancing_High-Resolution_Video-Language}), with none of them combining all the required modalities.
Similarly, ImageBind also makes use of multiple datasets (namely, AudioSet, SUN RGB-D \cite{SUN_RGB-D}, LLVIP \cite{LLVIP}, Ego4D \cite{Ego4D}, and \say{large-scale web datasets with (image, text) pairings}, that they seem to keep private), presumably due to a lack of simultaneous modalities and/or a small number of observations in each dataset.
Lastly, the NExT-GPT team curated their own public dataset (called MosIT), with all the modalities they were interested in, although it only encompasses $\mbox{5,000}$ observations.
We later compare the available datasets from the ones just mentioned with the one we generated.}

%% file: Publicacion/Estado_del_Arte.tex
\section{\blue{State of the Art}}\label{sec:Estado_del_Arte}

In the last decade, image generation has experienced enormous growth, driven by significant advances in fields such as artificial intelligence, machine learning and computer vision \cite{RenAIssance, The_Pile}. This progress has led to the creation of increasingly realistic and stylized images \cite{Image_Generation}.
While, thanks to advances in the quality of computer-generated images (with recent examples like Stable Diffusion XL \cite{SDXL} or 3 \cite{Stable_diffusion_3}, DALL·E 3 \cite{DALLE_3, Improving_Image_Generation_with_Better_Captions}, Imagen 3 \cite{Imagen3} or FLUX \cite{Flux}), the level of these images has reached a degree that makes it difficult to differentiate them from human-generated images; there is still much work to be done in terms of improving quality consistency, reducing bias, lowering computational costs, and facilitating user control over the generations (i.e. generating what the user actually expects/wants) \cite{Text-to-image_Diffusion_Models}.

\begin{figure*}[t]
\centering
\includegraphics[width=16.4cm]{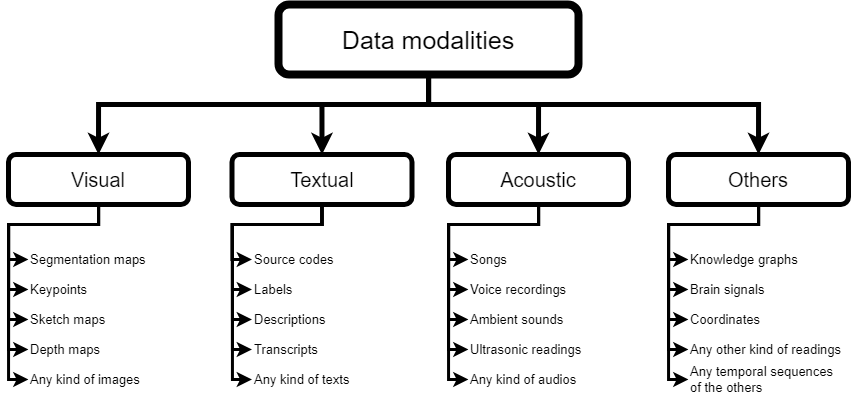}
\caption{Types of data modalities.}
\label{fig:modalidades}
\end{figure*}

To address this last challenge, one of the strategies that has been adopted is to increase the number of data modalities that the models receive (i.e. the types of data that are taken as input; e.g. text, image, audio, etc.) \blue{\cite{A_survey_of_multimodal_deep_generative_models, Any-to-Any_Generation, Multimodal_Learning_With_Transformers, gemini_1_5}}.
It is pertinent to comment that this increase in the number of modalities not only allows for greater control on the respective tasks, but also opens a way to perform new ones (for example, a detailed analysis can be seen in \cite{The_Dawn_of_LMMs}; where the capabilities of GPT-4V, a colossal multimodal model of text and images, are particularly studied).
In order to better illustrate the concept of data modalities, and inspired by the classification of data types explained in \cite{Multimodal_Image_Synthesis_and_Editing}, in Figure \ref{fig:modalidades} we present a conceptual map of the types of data modalities that can be used, along with examples for each.\footnote{\blue{For the sake of brevity, in our conceptual map we are just including the most popular examples.}}
An example of the use of multiple data modalities tends to be seen in image-to-image generation, where an image is taken as a reference to generate a new image, since the input image is usually accompanied by a text or a label to better condition/guide the final result \cite{Image-to-Image_Translation}.
In contrast, \blue{as seen in Section \ref{sec:Related_work},} audio conditioned image-to-image generation has not been explored as much as text conditioned image-to-image generation.
The latter may be because working with audio is not as intuitive as working with text \cite{MusicLM, Intuitive_Multilingual}, but that does not invalidate the potential benefit\blue{s} that could be obtained by using audio in certain scenarios (as those mentioned in Section \ref{sec:Introducción}).

Despite what we just said, we could still come up with ways to adapt the use of existing models to work with different data modalities than the ones that were originally intended for.
For instance, given the mentioned advancements in image-to-image models that are conditioned on textual inputs, it could be worth considering a new approach for scenarios where the objective is to perform image-to-image generations using audio instead of text.
A logical strategy for this goal could be to transcribe the audios into the corresponding textual representations/descriptions, which could then be utilized within existing text-image models.
This method should leverage the strengths of well established text-image models, potentially validating their use with audio-image data or of other kind, different to the originally intended text-image data.
However, it is crucial to acknowledge that, in addition to the fact that fields like audio-to-text conversion are still evolving and have not received as much attention as their visual counterparts \blue{\cite{Audio_Describing_Sound, Audio_Text_Models_Do_Not_Yet, Deep_Audio-visual_Learning, transcripter_generation}}, such approach presents several challenges that should be kept in mind.
Let us review the main ones:

\begin{enumerate}[label=\Alph*]
    \item Word limit in current models: currently, the problem of increasing the token window (i.e., words and characters) of text-to-image and audio-to-text models is open. For example, Stable Diffusion (an open-source neural network model that generates images based on text and/or image \cite{High-Resolution_Image_Synthesis}) has a context window of 75 tokens \cite{SD_Akashic_Records}.
    \item Compatibility between text-image and audio-text models: even if a capacity of hundreds of thousands of tokens is reached to describe any audio (as can be seen analogously in certain current text generation models \blue{\cite{The_Llama_3_Herd_of_Models, Mamba, Mistral_Models, Claude}}), the syntax of the text obtained with such an audio-text model must match that used by the respective text-image model with which it is to be combined, in order to maximize communication between the two \blue{\cite{High-Resolution_Image_Synthesis, The_Dawn_of_LMMs, Audio_Text_Models_Do_Not_Yet, Benchmarking_Cognitive_Biases}}.
    \item Noise incorporation\footnote{See \cite{What_is_noise} for a brief classical exploration of the definition of the term.}: in addition to the above, it has repeatedly been shown that transforming one modality to another is prone to incorporating noise or failing (to some extent) due to the noise that the data contains beforehand \blue{\cite{transcripter_generation, Nlip, Bicro, Noise_Aware_Learning}}. As a result, the more transformations we make, the more noise we risk adding in the process.
    \item Incorporation of biases: finally, it is pertinent to highlight that, influenced both by the data and their training architectures and configurations, models tend to prioritize and specialize in certain types of audio and have their own preferences for describing them \blue{\cite{Are_Models_Biased_on_Text, Word_Level_Explanations, On_Some_Biases_Encountered, Dont_Just_Assume}}. For example, typical cases of this can be seen in the underestimation/distortion of the order of events \cite{Audio_Text_Models_Do_Not_Yet, Benchmarking_Cognitive_Biases} or in the omission of details considered irrelevant \cite{Mirrorgan, Benchmarking_Cognitive_Biases}.
\end{enumerate}

It is due to these reasons that even if in some cases audios could/can be converted to texts and images conditioned with the generated texts, this is a significantly more problematic approach than just using audios and images.
For this reason, in this research we claim that, when working with a given set of modalities, it is convenient to perform the least number of data modality conversions possible. 
Furthermore, we believe that more audio-image research is needed to better address the respective tasks, instead of just trying to get by with what is already available.

Complementarily, it is relevant to point out that, as alluded to in \cite{I_Hear_Your_True_Colors, BLAT, BLAP, AudioSetCaps}, there are not many public datasets with audio-text pairs.
In our opinion, and despite the issues enumerated previously with modality transformation approaches, the best that can be done in this scenario is to leverage a model like CLIP \cite{CLIP} or BLIP \cite{BLIP}, which for a frame/image of a video could return a descriptive text. Said text could, in turn, be paired with a section of audio from the video that coincides with the time interval from which the frame was extracted.

The generation of audio-image-text observations could easily be automated, so the biggest challenge would lie in finding relevant and varied videos (as well as free from copyright conflicts). In any case, the videos collected in other research could be leveraged, within which there are recordings of musical instruments \cite{The_Sound_of_Pixels, MUSIC_dataset}, as well as various objects and animals \cite{Coordinated_Joint_Multimodal_Embeddings, AudioSetZSL}, and even different everyday environments \cite{SoundNet, YFCC100M}.

Regarding the kind of data collected, while it would be interesting to include relationships according to the lexical meaning of spoken words, as done in \cite{Audio-to-Visual_Cross-Modal} by relating spoken numbers and drawings of them, it would probably be better to focus on strictly non-abstract and non-artificial relationships (i.e., sounds only related to the recordings of when they were generated). This would restrict the training of the any model with this data (simplifying the range of relationships it must incorporate), facilitating convergence, and could even make its generations more intuitive.

In summary, we have noted a valuable opportunity to explore audio-image and audio-text tasks. This demands a great volume of data, for which there are not as many datasets as one would hope for nor there are common guidelines on how to collect it.
Due to this, in this research we precisely propose a method to obtain related audio-image-text observations from videos and we describe the datasets that we created with it.

%% file: Publicacion/Metodo.tex
\section{Videos to \blue{R}elated \blue{A}udio-\blue{I}mage-\blue{T}ext \blue{O}bservations}\label{sec:Metodo}

In the rapidly evolving landscape of multimodal data research, the integration of different kinds of data has become increasingly relevant.
In this this section, we will outline a systematic approach to generate audio-image-text observations from videos.
By leveraging high-quality video content, we aim to extract meaningful audio-image pairs and generate descriptive texts that enhance the utility of the resulting datasets.
This method should be helpful to face the current scarcity of comprehensive multimodal datasets.

\begin{figure*}[t]
\centering
\includegraphics[width=16.4cm]{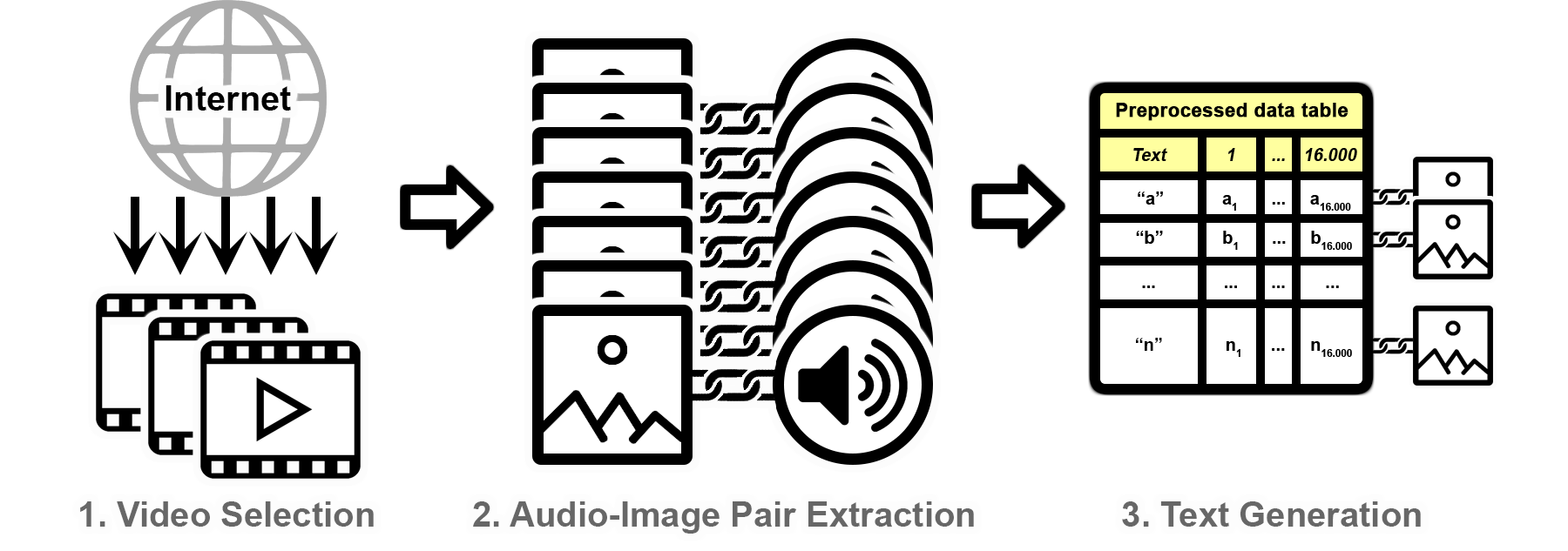}
\caption{Summary of the whole method. \textbf{1. Video Selection:} This initial phase involves identifying and selecting high-quality, continuous video recordings that feature synchronized audio and \blue{frames}, ideally ensuring a strong semantic connection between the modalities (i.e. both audio and image in each pair are extracted from and related to the same situation). \textbf{2. Audio-Image Pair Extraction:} In this step, audio segments are extracted from the selected videos, paired with corresponding frames, and processed to minimize file sizes while retaining essential information. \textbf{3. Text Generation:} The final phase utilizes an image-to-text model to generate descriptive texts for each audio-image pair, creating a comprehensive dataset with enhanced utility.}
\label{fig:resumen_metodo}
\end{figure*}

We will describe a multimodal data collection and processing method (specifically, for generating audio-image-text observations based on videos). It involves three key phases\blue{, which, for order sake, we will explain in subsections of their own}: 1. Initially, suitable videos are selected, prioritizing high-quality and continuous recordings, with synchronized audio and \blue{frames (see Subsection \ref{subsec:Video_selection})}. 2. The next phase involves extracting audio-image pairs from these videos, ensuring the audio is closely tied to the visual content and minimizing file sizes without significant information loss \blue{(see Subsection \ref{subsec:Video_selection})}. 3. Finally, the extracted pairs are used to generate descriptive texts using an image-to-text model, creating a comprehensive dataset for further use \blue{(see Subsection \ref{subsec:Text_generations})}.
\blue{All of this is} also illustrated in Figure \ref{fig:resumen_metodo}.

\blue{\subsection{Video Selection}\label{subsec:Video_selection}}

First of all, it is essential to talk about the videos that we would want to work with.
In addition to obviously avoiding copyrighted material and favoring HD videos with Hi-Fi audio, ideally we would hope to mainly use continuous recordings (i.e., without cuts or mixes), where each audio recording is strictly associated with the footage (i.e., without sounds that are not actually being produced in the images).
Ensuring that the audio is strictly associated with the corresponding images/frames will allow for a consistent and accurate semantic connection between them, regardless of the task for which the data is being used. Additionally, using continuous recordings increases the likelihood of finding suitable video fragments to convert, especially when seeking longer audio segments.

Once again, as said in Section \ref{sec:Estado_del_Arte}, one can leverage public material from other research, like that from \cite{Awesome-Video-Datasets}.
After we have collected our videos, we can start extracting pairs that consist of an image and its corresponding audio.

\blue{\subsection{Audio-Image Pair Extraction}\label{subsec:Audio-image_pair_Eetraction}}

Let us define the properties of the images and audios with which we will work, designed to minimize their size as much as possible, while preserving their core contents.
Based on our own experience and on what has been seen in other works that generated good results \cite{Listen_to_Look, Audio-to-Visual_Cross-Modal, Audio_deepfakes, TimbreCLIP}, we would generally advise for the images to have a resolution of 512x512 pixels, in .jpg or .png format (.jpg is probably the best option, as it usually uses less space) and in RGB24 (a standard color model, consisting of a red channel, a green channel, and a blue channel, with values ranging from 0 to 255); while for the audios we would suggest a duration of 1 \sixs{}, with 16,000 \sixhz{}, 16 \sixbits{} depth, in .wav format and monophonic (i.e., with a single channel). In fact, these are the properties we chose for the datasets that we will show in Section \ref{sec:Datasets}.

\begin{figure*}[t]
\centering
\includegraphics[width=16.4cm]{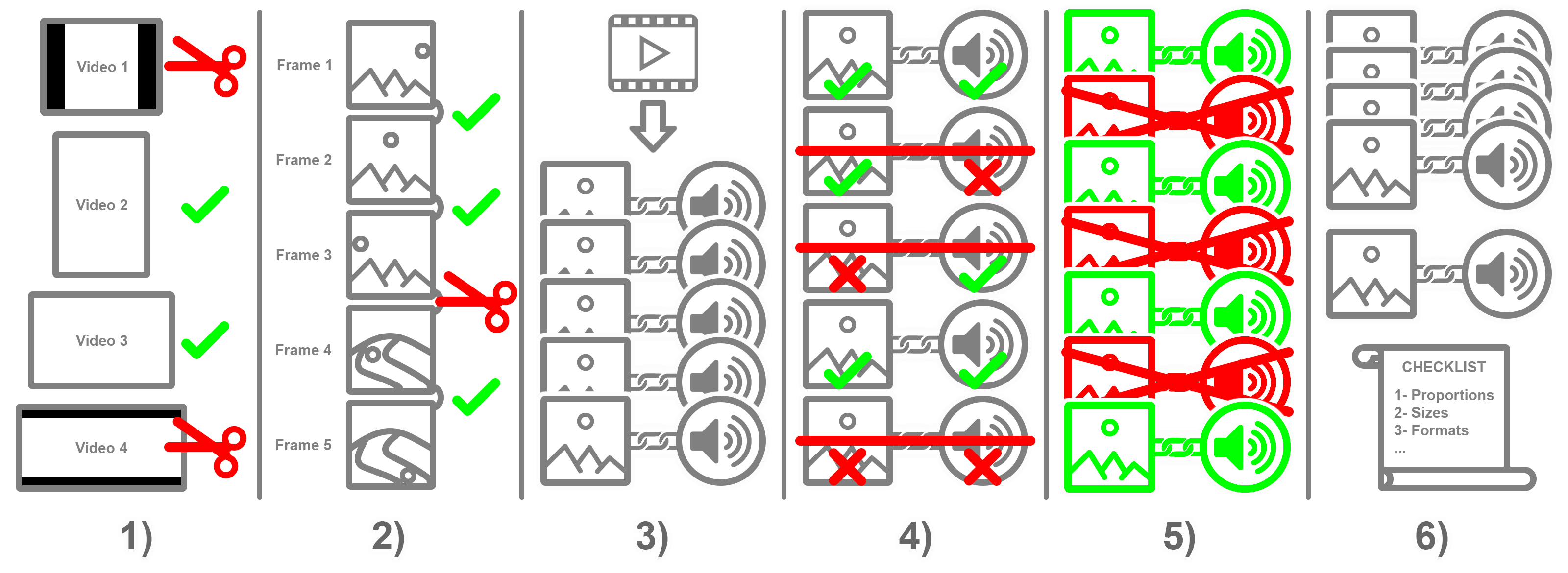}
\caption{Summary of the audio-image pair extraction procedure. 1) Removal of black borders. 2) Discontinuous footage separation. 3) Initial audio-image pair extraction. 4) Discard of deficient audio-image pairs. 5) Skipping of pairs to enhance diversity. 6) Enforcement of the correct properties.}
\label{fig:resumen_extraccion_pares}
\end{figure*}

Regarding the strategy for extracting audio-image pairs, the following procedure is proposed (which is also summarized in Figure \ref{fig:resumen_extraccion_pares}):
\begin{enumerate}
    \item Inspect each video, evaluating if it has black borders. If so, these must be cropped to only consider relevant information in the final images.
    This can be accomplished by taking a frame in the middle, and verifying if each first and last column/row does not have a pixel with value higher than a certain threshold in any of its channels (we use a threshold of 15 for this).
    In that situation, that column/row should be deleted and the step is repeated until all of the black borders have been erased \blue{(similar to what is done in \cite{Phantom-Data})}.
    \item To ensure that no drastic/unnatural changes are present in any audio, go through each video frame by frame.
    If an abrupt change is detected (for example, if the average of the squared differences of pixels between two consecutive frames is greater than a threshold of 90), then proceed to divide the footage into two and, for all purposes, treat them as distinct videos going forward.
    It should be \blue{remarked} that the videos with fade transitions could present some problem with this approach and, to compensate it, more future frames could be used in the comparison.
    \item For each resulting video fragment, extract consecutive audios of one second, along with the frame that is approximately in the middle of that time interval to form the respective pairs.
    \blue{Please note that this is known as middle frame extraction, and it is a well-extended heuristic to select a representative frame of a video fragment, which should have better odds to properly match semantically with the respective audio \cite{A_Comparison_Between_KeyFrame_Extraction, Key_Frame_Selection_for_Temporal_Graph, ActionAtlas, A_Structured_Model_for_Action_Detection}.}
    If the final part does not reach one second, it must be ignored.
    \item Discard pairs whose audio has at least a given amount of continuous silence, as they will probably not contain enough information to be useful (we looked for continuous intervals of 0.5 \sixs{} where none of their samples had an absolute value higher than 100).\footnote{Keep in mind that we are considering samples of 16 \sixbits{}, implying that the values they take go from -32,768 up to 32,767.}
    This, in turn, can be combined with a discount of pairs where the mean of all pixels in the image does not surpass a given threshold (we suggest a threshold of 10).
    The latter should further ensure that no frames too dark are included.
    \item To increase diversity in the data (and thus not skew the research), also consider skipping a given number of pairs from each video fragment (we just kept one pair from every three).
    \item To preserve the dimensions, crop each frame according to the smaller dimension and around the center of the image, and rescale to 512x512 pixels. In addition, make sure to use the correct configuration for both saved files.
\end{enumerate}

\begin{figure*}[t]
\centering
\includegraphics[width=13.4cm]{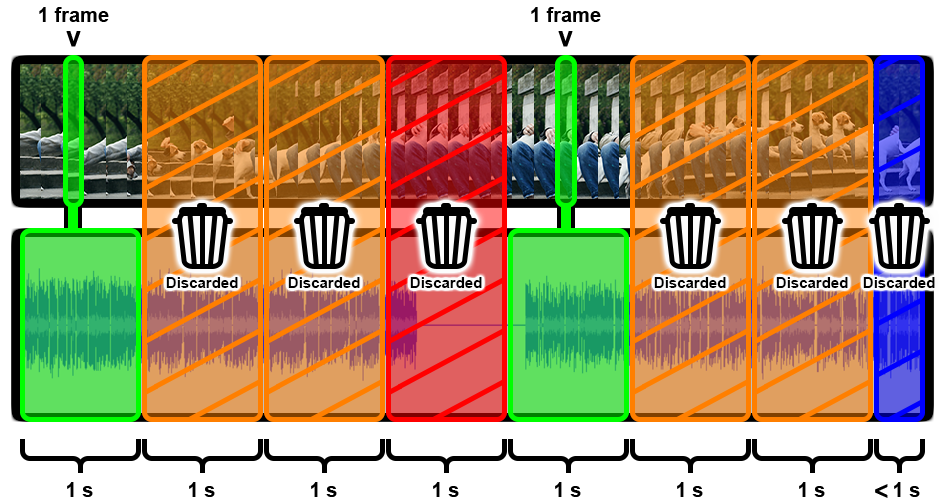}
\caption{Application example of steps 3, 4 and 5 of the audio-image pairs generation process, where the pairs are extracted from a video fragment and filtered according to our needs.}
\label{fig:extracción_de_pares}
\end{figure*}

An example of steps 3, 4 and 5 of this process, where the extraction and filtering of audio-image pairs is used on an isolated video fragment, is shown in Figure \ref{fig:extracción_de_pares}.
The blue fragment is discarded, as it is the last one and it does not reach a duration of 1 \sixs{}.
The red fragment is also not considered, due to having at least 0.5 \sixs{} of continous silence.
And, finally, just one from every three pairs is considered (denoted by their green color), in order to increase the diversity in the data.

\begin{figure*}[t]
\centering
\includegraphics[width=16.4cm]{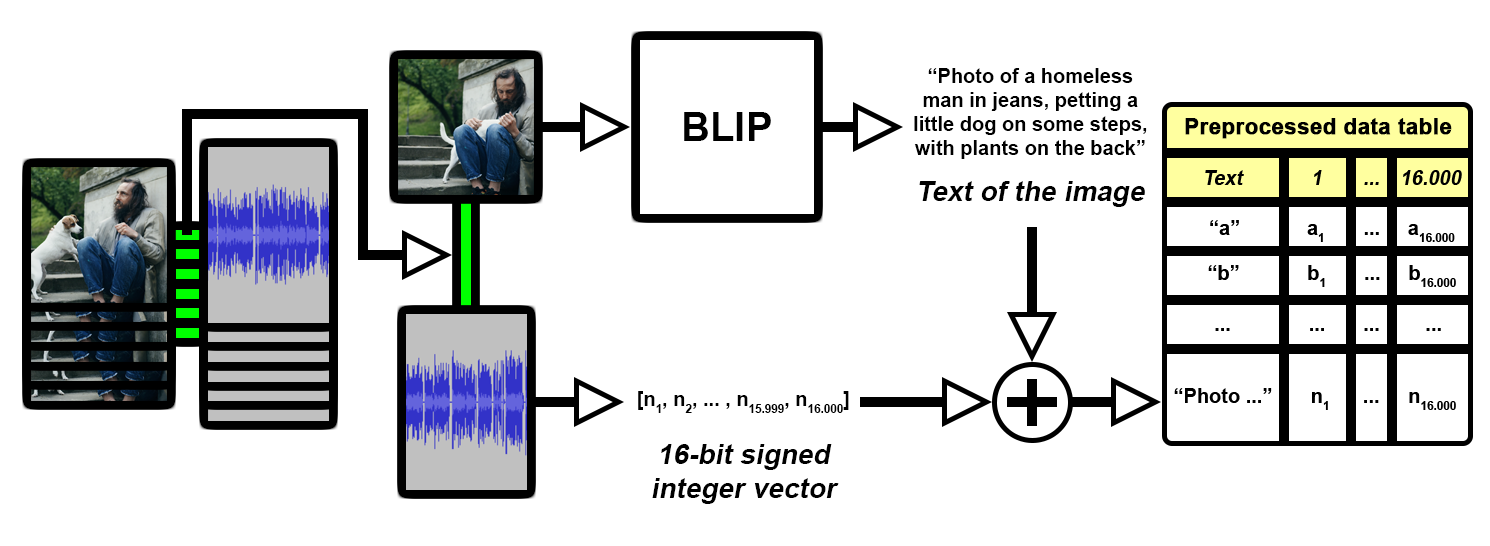}
\caption{Example of the final data preprocessing. Audio-image pairs are expanded to include a textual modality, by generating descriptive texts based on each image. These texts, along with corresponding audio values, are saved in a structured table to ease the use of the resulting dataset.}
\label{fig:preprocesamiento}
\end{figure*}

\blue{\subsection{Text Generations}\label{subsec:Text_generations}}

After we generated our audio-image pairs, we are ready to create the respective texts for each one of them.
As \blue{suggested} in Section \ref{sec:Estado_del_Arte}, this will be accomplished by taking each image from every pair and, based on it, generating an associated text with the image-to-text model BLIP \cite{BLIP}\footnote{For the sake of speed, we used 2 beams, with a minimum length of 10 tokens and a maximum of 20.}; while the audio will be represented as a vector of 16,000 elements, with signed 16 \sixbits{} integers.
\blue{The motivation of using an image-to-text model lies in the fact that the manual writing of textual descriptions for each audio-image pair is a time consuming process that makes it impractical for a large number of observations.
It is worth commenting that this modern possibility of leveraging image-to-text models is not something particularly novel and has also been validated in similar research \cite{OpenHumanVid, Learning_Text-to-Video_Retrieval_from_Image_Captioning, UltraVideo}.
A reasonable alternative would be to employ an audio-to-text model instead (like the ones mentioned in Table \ref{tab:resumen_investigaciones}), although such models still need more development before being used reliably in tasks like this.}
In the end, both the text and the vector will then be added as a new observation/entity in a data table, so they can be manipulated more easily.
For better results, it should also be pointed out that if the reader has access to a more advanced computer, then a more sophisticated image-to-text model (like BLIP-2 \cite{BLIP-2}) should be leveraged instead of BLIP.

An illustration of how this preprocessing would look like is shown in Figure \ref{fig:preprocesamiento}.
It is \blue{relevant to note} that the use of a table to save the resulting data is an optional step and the data can be stored in any form that best suits the user.

An interesting nuance to highlight is that, in the world of audio processing, there has been a tendency to prefer converting audios to spectrograms, moving from the temporal space to the temporal-spectral space, in order to facilitate pattern extraction with classical methods.
This is still seen with more modern techniques \cite{Listen_to_Look, Audio-to-Visual_Cross-Modal, Wav2CLIP}, but, in this paper, we are only interested in creating datasets with common audio.
Therefore, such conversion is omitted in our case.

It is also worth mentioning that the minimum number of observations (or samples) \say{necessary} to train machine learning models is a topic open to debate (even for LLMs \cite{Training_Compute_Optimal_LLM}).
While a popular rule of thumb is to employ at least ten times the number of parameters of the respective model, more formal and older estimations determined that twenty times the number of parameters would be reasonable \cite{Revisiting_Sample_Size}.
This aspect should be kept in mind when generating any dataset to train machine learning models.

Lastly, we can comment that one could even incorporate complementary data, created by any-to-any models \cite{Any-to-Any_Generation, Codi2, Imagebind}.
While this could be enticing at a first glance, it is a must to always remember the concerns presented in Section \ref{sec:Estado_del_Arte}, about generating or converting data with third party models (whether publicly validated or not).
For most cases, we strongly advise \blue{prioritizing} non-artificial data.

%% file: Publicacion/Dataset.tex
\section{Results and Discussion}\label{sec:Datasets}

\begin{figure*}[t]
\centering
\includegraphics[width=16.4cm]{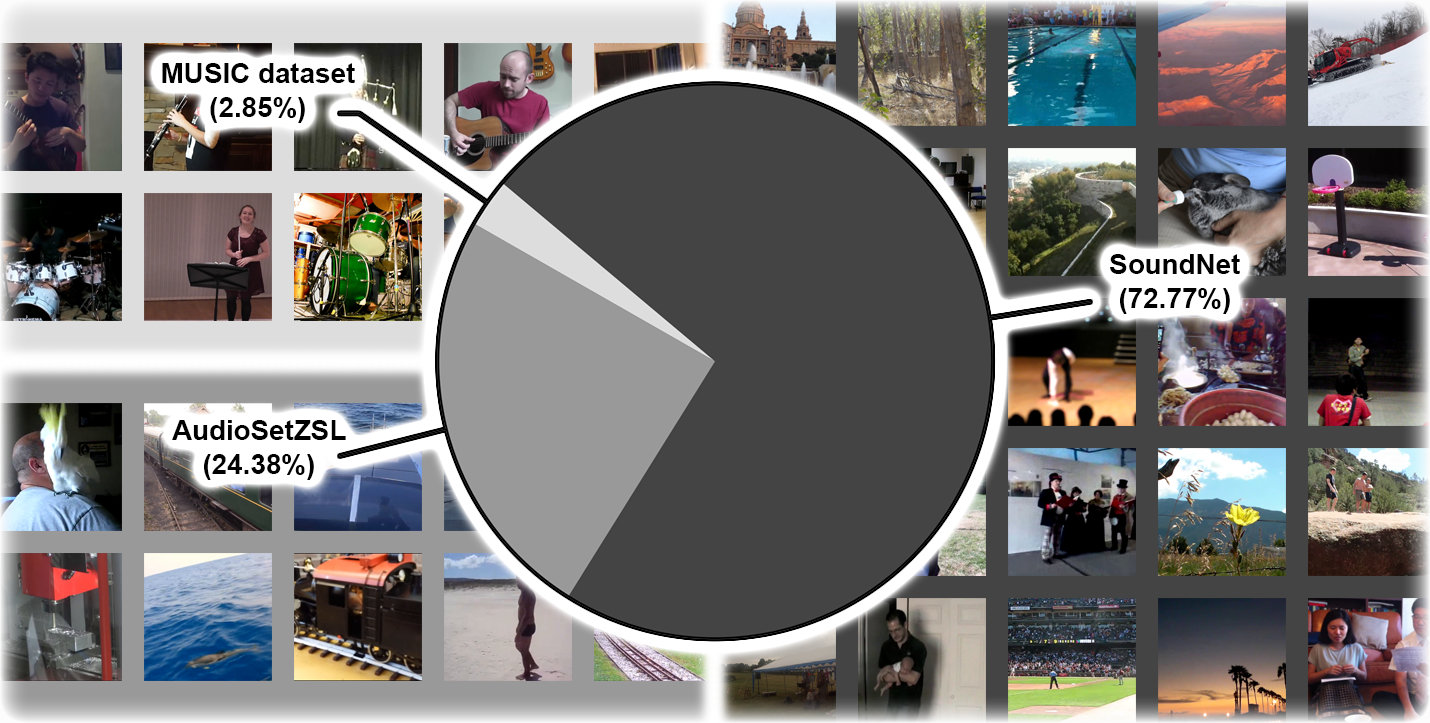}
\caption{\blue{Visualization of the sources of our data, with the approximate percentage for each dataset and some real examples resulting from each one.}}
\label{fig:conteo_muestras}
\end{figure*}

This section presents the datasets we created using the method described in Section \ref{sec:Metodo}.

To create our audio-image pairs, we utilized videos from the public datasets MUSIC dataset \cite{The_Sound_of_Pixels, MUSIC_dataset}, AudioSetZSL \cite{Coordinated_Joint_Multimodal_Embeddings, AudioSetZSL}, and SoundNet \cite{SoundNet, YFCC100M}.
\blue{The videos we employed from MUSIC dataset only contain solo performances of twenty-one different kinds of instruments, while the other two are much more diverse, ranging from musical instruments, to various objects and animals, and different everyday environments.
This diversity is relevant, as it brings substantial versatility to our final dataset.}
Given that AudioSetZSL is intended solely for research purposes, our datasets will also be made available for research use only, ensuring that they contribute to the advancement of multimodal data analysis while adhering to ethical standards in data sharing.

\begin{figure*}[t]
\centering
\includegraphics[width=16.4cm]{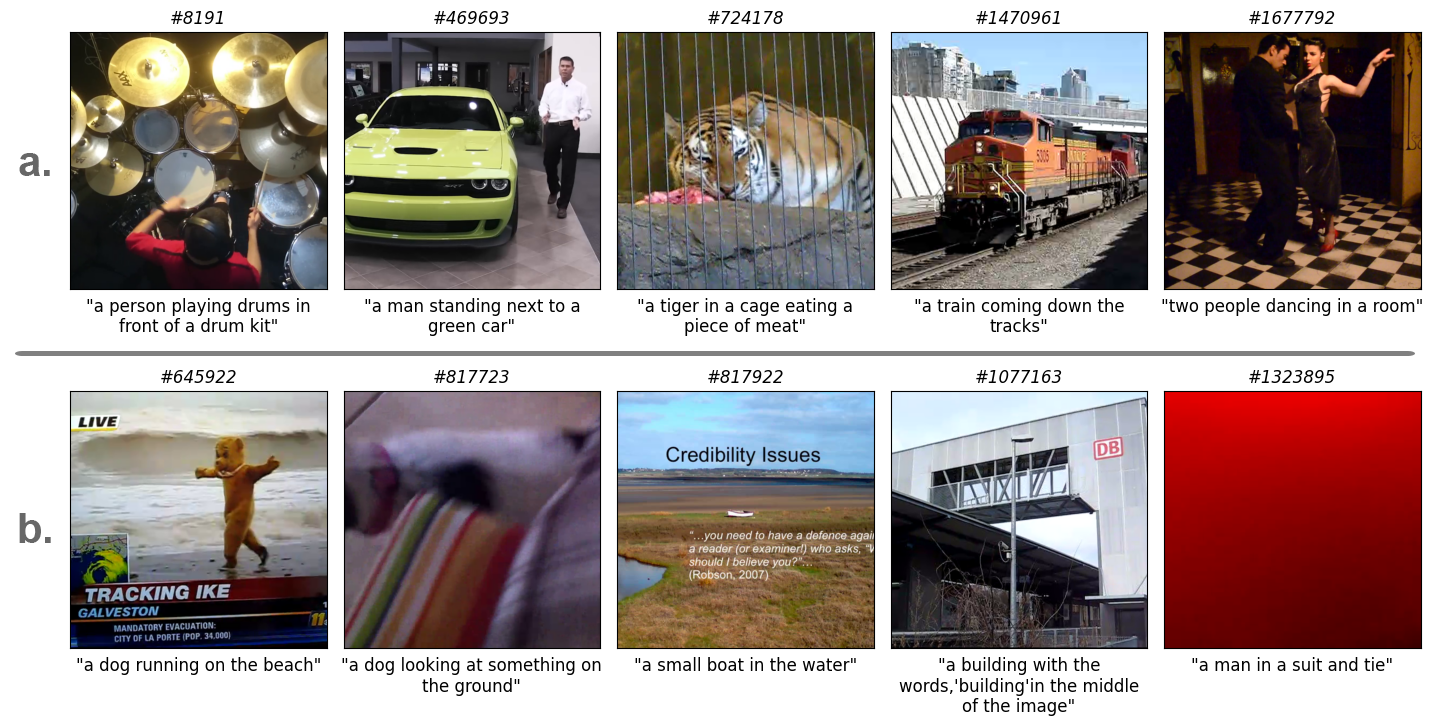}
\caption{(a.) A small selection of what we label sufficient and (b.) insufficient quality image-text pairs, from a random sample of 60 observations (all of them available with their respective audios at \cite{AVT_Multimodal_Dataset}). We deem \textit{\#645922} insufficient because the image has text and is from a screen; while the associated text is wrong on subject of the sentence, as it clearly shows a person in a costume and not a dog. \textit{\#817723} is insufficient as the image is too blurry to make a reasonable guess on what it is showing. \textit{\#817922} has text in the image and the associated text is wrong. In \textit{\#1077163} the text is also mistaken. Finally, \textit{\#1323895} has an useless image and a made up description.}
\label{fig:muestra_imagenes}
\end{figure*}

We applied the method outlined in Section \ref{sec:Metodo} to process 282,081 videos, resulting in the generation of 2,240,231 audio-image pairs. 
\blue{From these pairs,} $\mbox{63,849}$ come from MUSIC dataset, $\mbox{546,254}$ from AudioSetZSL and $\mbox{1,630,128}$ from SoundNet \blue{(proportions that can be further appreciated in Figure \ref{fig:conteo_muestras})}.
These pairs have been named with numbers, in a rising manner, and organized into 639 separate .zip files, which are publicly accessible on Kaggle, categorized into three distinct datasets \cite{Pares1de3, Pares2de3, Pares3de3}. This structured approach not only facilitate\blue{s} easy access for researchers, but also promote\blue{s} further exploration and utilization of the datasets in various multimodal applications.

It is relevant to point \blue{out} that some possibly problematic frames for certain uses were deemed acceptable by our filters. Namely, we noted that frames with text, with blurry images and/or with mainly a plain color were included (see images \textit{\#817922}, \textit{\#817723} and \textit{\#1323895}, respectively, from row b. in Figure \ref{fig:muestra_imagenes}). This reinforces the value of corroborating that the original videos selected for the audio-image pair extraction process align with our interests, meaning that it would be ideal to make sure that no video with flaws that we cannot fix should be considered in the first place.
Of course, curating lists of hundreds of thousands of videos is unfeasible for many researchers, which implies that the heavy work must focus on harnessing videos collected by other individuals, as well as applying the respective filters to assure the desired properties.
As seen with our results, the specific filters we employed seem to still have room for improvement.
In any case, these cases we mentioned are a minority in our data. Nevertheless, it is important to keep this in mind and we think that the removal of these pairs could also lead to some interesting research.

\blue{In a subsequent step}, we generated descriptive captions for all of the images, using the BLIP model. We paired these texts with the respective audios and stored them in 893 .csv tables. These, in turn, were saved in 263 .zip files, along with the associated images and preserving the numeric names. The final audio-image-text data can be found in 4 public datasets that we uploaded to Kaggle \cite{Pares1de4, Pares2de4, Pares3de4, Pares4de4}.
In addition, as we cannot properly share audio through this document, we have prepared a public page, where we share 60 random samples of our final datasets, to give a more solid idea of our results \cite{AVT_Multimodal_Dataset}.

\begin{table}[t]
    \centering
    \begin{tabular}{p{4.1cm}|p{3.5cm}|p{3.65cm}|p{3.5cm}}
    \#1 people:\hfill 15.67\% & \#16 white:\hfill 3.84\% & \#31 train:\hfill 2.55\% & \#46 red:\hfill 1.56\% \\ 
    \#2 man:\hfill 15.58\% & \#17 road:\hfill 3.68\% & \#32 sky:\hfill 2.48\% & \#47 beach:\hfill 1.55\% \\ 
    \#3 person:\hfill 9.31\% & \#18 words:\hfill 3.44\% & \#33 building:\hfill 2.39\% & \#48 tree:\hfill 1.51\% \\ 
    \#4 group:\hfill 9.14\% & \#19 two:\hfill 3.35\% & \#34 floor:\hfill 2.32\% & \#49 bird:\hfill 1.4\% \\ 
    \#5 car:\hfill 8.93\% & \#20 driving:\hfill 3.21\% & \#35 dog:\hfill 2.24\% & \#50 guitar:\hfill 1.38\% \\ 
    \#6 playing:\hfill 7.66\% & \#21 table:\hfill 3.18\% & \#36 middle:\hfill 2.01\% & \#51 shirt:\hfill 1.36\% \\ 
    \#7 sitting:\hfill 7.65\% & \#22 crowd:\hfill 3.14\% & \#37 cars:\hfill 1.94\% & \#52 boat:\hfill 1.35\% \\ 
    \#8 room:\hfill 7.25\% & \#23 suit:\hfill 3.13\% & \#38 holding:\hfill 1.77\% & \#53 parking:\hfill 1.35\% \\ 
    \#9 street:\hfill 6.83\% & \#24 field:\hfill 2.99\% & \#39 child:\hfill 1.7\% & \#54 little:\hfill 1.27\% \\ 
    \#10 background:\hfill 6.6\% & \#25 water:\hfill 2.97\% & \#40 trees:\hfill 1.69\% & \#55 band:\hfill 1.27\% \\ 
    \#11 down:\hfill 5.8\% & \#26 front:\hfill 2.96\% & \#41 riding:\hfill 1.67\% & \#56 girl:\hfill 1.25\% \\ 
    \#12 standing:\hfill 4.68\% & \#27 city:\hfill 2.93\% & \#42 cat:\hfill 1.67\% & \#57 truck:\hfill 1.25\% \\ 
    \#13 woman:\hfill 4.58\% & \#28 tie:\hfill 2.87\% & \#43 laying:\hfill 1.66\% & \#58 bed:\hfill 1.23\% \\ 
    \#14 walking:\hfill 4.44\% & \#29 parked:\hfill 2.73\% & \#44 black:\hfill 1.61\% & \#59 chair:\hfill 1.19\% \\ 
    \#15 baby:\hfill 3.93\% & \#30 stage:\hfill 2.66\% & \#45 night:\hfill 1.56\% & \#60 wall:\hfill 1.16\% \\ 
    \end{tabular}
    \caption{The top 60 words that appear in most observations of our final datasets. Prepositions, pronouns, conjunctions and determiners are not considered, and percentages in parenthesis show the proportion of observations that include them.}
    \label{tab:top_palabras}
\end{table}

Going more into detail regarding the final data, we conducted a small statistical study across all the texts.
We confirmed that all descriptions have a length from 1 up to 16 words, where the mean is $7.37$ and the standard deviation is $1.74$, approximately\blue{. We regard these} values as appropriate to avoid redundancies from the image-to-text model.
\blue{For comparison, we can comment that the well-known acoustic-textual dataset AudioCaps ended up with an average of $9.03$ words per description \cite{AudioCaps}, which does not stray too far from our result.}
We also counted the number of different words in the texts and found out that there were $\mbox{8,824}$ different words in use.
From the list of different words, we discarded prepositions, pronouns, conjunctions and determiners, ending up with a new total of $\mbox{8,652}$ different words.
Finally, we went through the latter preprocessed list, counting the number of times that each word appeared in an observation (counting just once per observation).
We show the top 60 words with the biggest percentages of presence across all observations in Table \ref{tab:top_palabras}.
From this, we can confirm that, despite a significant amount of observations containing situations featuring people, these are not the majority according to the text descriptions.
Moreover, as planned, there is a nice range of diversity, given the varied collection of words that can be seen in Table \ref{tab:top_palabras}.

\begin{figure*}[t]
\centering
\includegraphics[width=16.4cm]{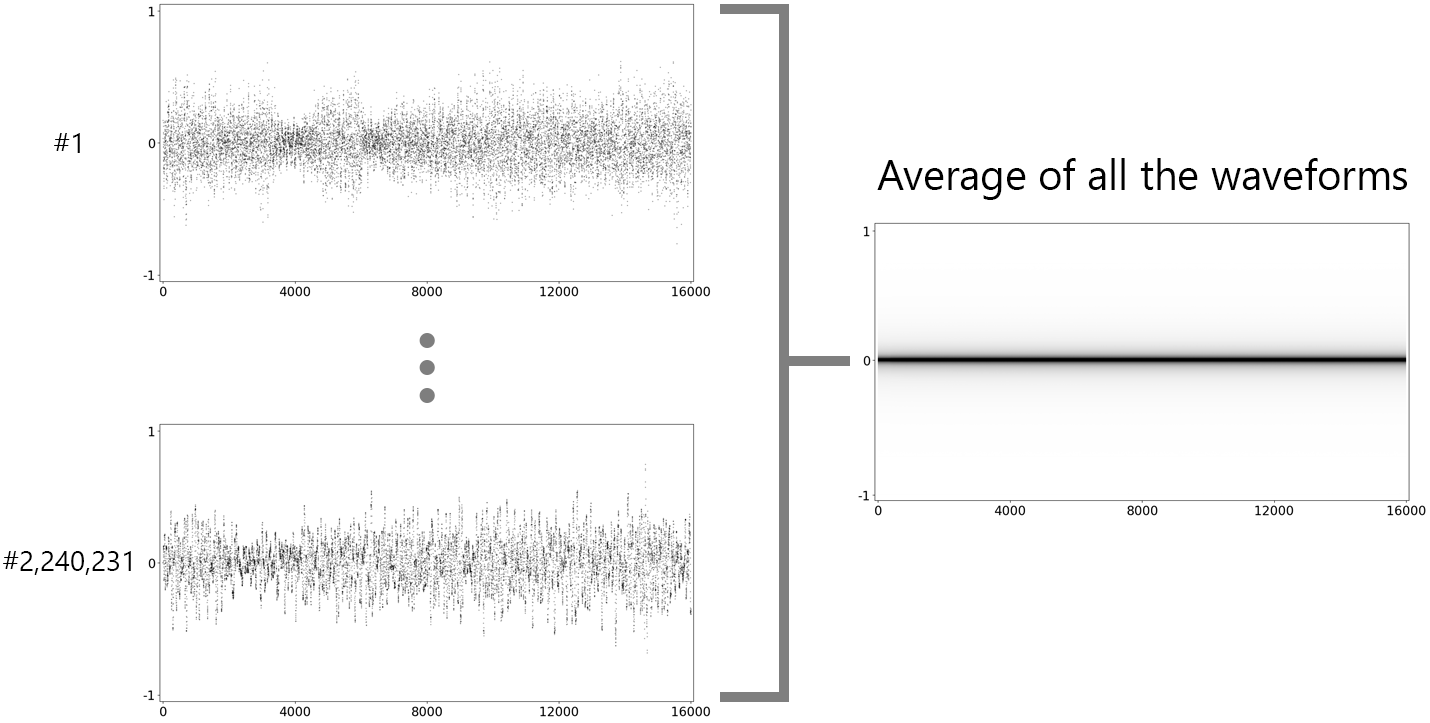}
\caption{\blue{Average of all the waveforms in our observations. The horizontal axis contains the timestamps, while the vertical one is for the instantaneous amplitudes.}}
\label{fig:average_waveform}
\end{figure*}

\begin{figure*}[t]
\centering
\includegraphics[width=16.4cm]{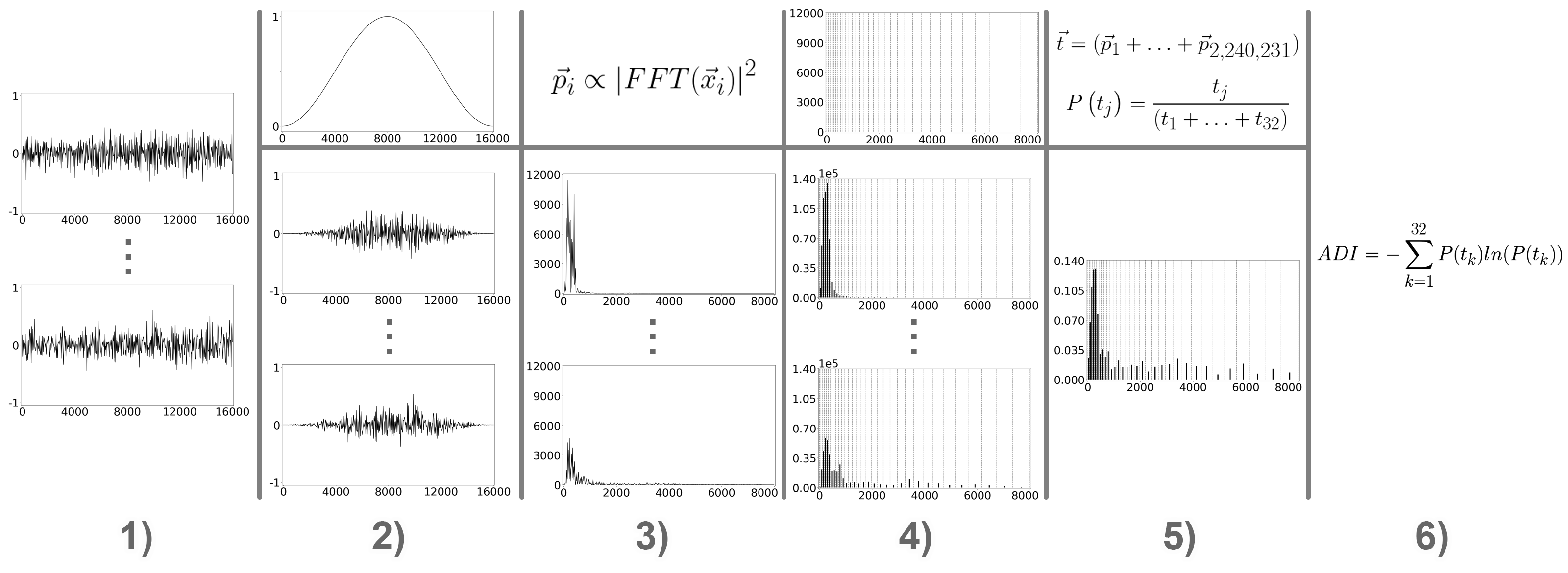}
\caption{\blue{Summary of our ADI calculation. 1) We take all of our audios in their raw form. 2) We apply the Hann function \cite{hann_window} (visible on the top) over each audio signal, so all of them loop smoothly and we avoid spectral leakage. 3) We compute the fast Fourier transform of each signal from the previous step, we get the magnitude of each resulting complex number and square them; now these new values are proportional to the real power spectrums, and we can treat these as substitutes of them in the next steps. 4) We generate 32 evenly spaced bins in the mel scale \cite{mel_scale} of our range of frequencies (i.e. $[0, \mbox{8,000}]$) and aggregate the respective values that share each bin. 5) We sum all of our 2,240,231 vectors of grouped power spectrums, preserving their bins and dividing each resulting component by the sum of all of them combined; this effectively leaves us with the probabilities of presence of each interval of frequencies in our audios. 6) Finally, we apply the Shannon index \cite{shannon_index} over our probabilities of the previous step, in order to obtain our ADI.}}
\label{fig:ADI}
\end{figure*}

\blue{To attest to the usefulness of our data, we also conducted two additional tests to inspect both biases and diversity in our audios.
On one hand, for the biases, we created a $\mbox{65,536}\!\times\!\mbox{16,000}$-matrix (coinciding with our chosen bit depth and total samples per audio, respectively), filled with zeros, and proceeded to add to each element the count of times where the corresponding instantaneous amplitude was present in the given timestamps, across all the audios.
We then plotted the resulting matrix (assigning to zero the white color and to the maximum count of the matrix a the black color, with all the counts in between a grey that linearly denotes its closeness to each extreme) and obtained the result on the right of Figure \ref{fig:average_waveform}, which also illustrates the whole procedure. 
The observable Gaussian distributions in all the timestamps coincide neatly with the theory \cite{Does_mixing_of_speech_signals_comply, Central_Limit_Theorem}, and thus this shows that no evident biases are present.
On the other hand, for the diversity, we calculated the corresponding Acoustic Diversity Index (ADI) \cite{Soundscape_diversity, Using_acoustic_indices_in_ecology, Soundscape_Ecology, The_Acoustic_Index}.
This is a popular metric, based on the Shannon index \cite{shannon_index}, and with high popularity for audio diversity measurements (especially in fields related to biology, although it can be employed with any kind of audio dataset).
We summarized its calculation on Figure \ref{fig:ADI}, and we also need to point out that, in our case, this metric may take values between 0 and $\sim 3.4657$ (with bigger values conveying a greater diversity).
Dividing this interval into three equally distributed ones, we end up with the following: $[0, 1.1552]$ for low diversity values, $(1.1552, 2.3105)$ for medium diversity values, and $[2.3105, 3.4657]$ for high diversity values.
Our resulting ADI was $\sim 3.0525$, which serves as complementary evidence of the diversity in our dataset.}

\begin{table}[!t]
    \footnotesize
    \begin{tabular}{|m{0.2\textwidth}|m{0.02\textwidth}|m{0.02\textwidth}|m{0.02\textwidth}|m{0.02\textwidth}|m{0.1\textwidth}|m{0.3\textwidth}m{0.0\textwidth}|}
        \hline
        \centering \textbf{Name of the dataset} & \centering \textbf{A} & \centering \textbf{I} & \centering \textbf{T} & \centering \textbf{V} & \centering \textbf{\# of samples} & \multicolumn{2}{m{0.435\textwidth}|}{\centering \textbf{Contents}} \\
        \hline
        AudioCaps \cite{AudioCaps} & \centering \yesj & \centering \noj & \centering \yesj & \centering \noj & \centering $>45.5$K & \multicolumn{2}{m{0.435\textwidth}|}{Alarms, various objects and animals, natural phenomena, and different everyday environments.} \\
        \hline
        AudioSet \cite{AudioSet} & \centering \yesj & \centering \noj & \centering \yesj & \centering \noj & \centering $>2.0$M & \multicolumn{2}{m{0.435\textwidth}|}{632 audio event classes, including musical instruments, various objects and animals, and different everyday environments.} \\
        \hline
        CMU-MOSEI \cite{CMU-MOSEI} & \centering \yesj & \centering \yesj & \centering \yesj & \centering \noj & \centering $>3.2$K & \multicolumn{2}{m{0.435\textwidth}|}{People speaking directly to a camera in monologue form, intended for sentiment analysis.} \\
        \hline
        Ego4D \cite{Ego4D} & \centering \yesj & \centering \noj & \centering \noj & \centering \yesj & \centering $>5.8$K & \multicolumn{2}{m{0.435\textwidth}|}{Egocentric video footage of different everyday situations, with portions of the videos accompanied by audio and/or 3D meshes of the environment.} \\
        \hline
        Flickr30k Entities \cite{Flickr30k_Entities} & \centering \noj & \centering \yesj & \centering \yesj & \centering \noj & \centering $>31.7$K & \multicolumn{2}{m{0.435\textwidth}|}{Diverse environments, objects, animals, and activities, with the addition to bounding boxes to the image-text pairs.} \\
        \hline
        Freesound 500K \cite{Any-to-Any_Generation} & \centering \yesj & \centering \noj & \centering \yesj & \centering \noj & \centering $500.0$K & \multicolumn{2}{m{0.435\textwidth}|}{Diverse situations, sampled from the Freesound website, and accompanied by tags and descriptions.} \\
        \hline
        HD-VILA-100M \cite{Advancing_High-Resolution_Video-Language} & \centering \noj & \centering \noj & \centering \yesj & \centering \yesj & \centering $>100.0$M & \multicolumn{2}{m{0.435\textwidth}|}{A wide range of categories, including tutorials, vlogs, sports, events, animals, and films.} \\
        \hline
        InternVid \cite{InternVid} & \centering \noj & \centering \noj & \centering \yesj & \centering \yesj & \centering $>233.0$M & \multicolumn{2}{m{0.435\textwidth}|}{Diverse environments, objects, animals, activities, and everyday situations.} \\
        \hline
        LAION-400M \cite{LAION-400M} & \centering \noj & \centering \yesj & \centering \yesj & \centering \noj & \centering $400.0$M & \multicolumn{2}{m{0.435\textwidth}|}{Everyday scenes, animals, activities, art, scientific imagery and various objects.} \\
        \hline
        LLVIP \cite{LLVIP} & \centering \noj & \centering \yesj & \centering \noj & \centering \noj & \centering $>16.8$K & \multicolumn{2}{m{0.435\textwidth}|}{Street environments, where each visible light image is paired with an infrared one of the same scene.} \\
        \hline
        MMIS \cite{MMIS} & \centering \yesj & \centering \yesj & \centering \yesj & \centering \noj & \centering $>150.0$K & \multicolumn{2}{m{0.435\textwidth}|}{A wide range of interior spaces, capturing various styles, layouts, and furnishings.} \\
        \hline
        MosIT \cite{NExT_GPT} & \centering \yesj & \centering \yesj & \centering \yesj & \centering \yesj & \centering $5.0$K & \multicolumn{2}{m{0.435\textwidth}|}{Diverse environments, objects, animals, artistic elements, activities, and conversations.} \\
        \hline
        SUN RGB-D \cite{SUN_RGB-D} & \centering \noj & \centering \yesj & \centering \noj & \centering \noj & \centering $>10.0$K & \multicolumn{2}{m{0.435\textwidth}|}{Everyday environments, where each image has the depth information of the various objects in it.} \\
        \hline
        SoundNet \cite{SoundNet} & \centering \yesj & \centering \noj & \centering \noj & \centering \yesj & \centering $>2.1$M & \multicolumn{2}{m{0.435\textwidth}|}{Videos without professional edition, depicting natural environments, everyday situations, and various objects and animals.} \\
        \hline
        WebVid-10M \cite{Frozen_in_Time} & \centering \noj & \centering \noj & \centering \yesj & \centering \yesj & \centering $>10.0$M & \multicolumn{2}{m{0.435\textwidth}|}{Natural environments, everyday situations, and various objects and animals.} \\
        \hline
        \textbf{AVT Multimodal Dataset (Ours)} & \centering \yesj & \centering \yesj & \centering \yesj & \centering \noj & \centering $>2.2$M & \multicolumn{2}{m{0.435\textwidth}|}{Musical instruments, various objects and animals, and different everyday environments.} \\
        \hline
    \end{tabular}
    \caption{\blue{A comparison table between many multimodal datasets and ours. \textbf{A} means that the observations include \textbf{Audios}, \textbf{I} means the same for \textbf{Images}, \textbf{T} for \textbf{Texts} and \textbf{V} for \textbf{Videos}. \yesj{} means the data modality is present in the respective dataset. K stands for thousands and M for millions.}}
    \label{tab:comparacion_datasets}
\end{table}

\blue{Now, to offer a clearer reference of the contribution of our dataset, pay attention to Table \ref{tab:comparacion_datasets}, where we compare our dataset with the ones mentioned in Section \ref{sec:Related_work} (leveraged by models that include acoustic, textual and visual modalities) and some additional ones that could also be employed in similar audio-image-text tasks.
As we can see, most datasets do not even contain one million observations, which is a real handicap, given that modern models deal with millions of parameters and therefore require larger datasets to be properly trained.
Currently, researchers need to arduousness search for multiple datasets and artfully come up with ways to utilize them in audio-image-text tasks; as they not only be too small, but do not contain all the modalities needed and/or their contents are too specialized (not to mention the extra preprocessing steps one must add when the data is not homogeneous).
All of this hinders the potential research that could be done in the field, and thus we expect both our dataset and our detailed method contribute to ease this struggle, especially when noticing the high supply of audiovisual datasets.}

Once again, there is a shortcoming that we must highlight.
Despite the relatively long time taken to create the text descriptions, the BLIP configuration used was fairly basic to maximize speed.
This means that the text quality is not nearly as high as one would wish for in some instances.
To illustrate the latter, let us look at Figure \ref{fig:muestra_imagenes}.
Contrasting with the appropriate descriptions we get in cases like row a., row b. presents a diverse kind of errors. \textit{\#645922} misidentifies a person in a costume as a dog, \textit{\#817723} has an imprecise caption due to the poor resolution of the image, \textit{\#817922} may also be negatively affected by the presence of texts, \textit{\#1077163} straight up imagines a text that does not exist, and \textit{\#1323895} hallucinates the presence of a man when it is clearly just a color gradient.
Again, the quality of the images we work with has a fundamental influence in the quality of our final texts, but so does the model we use.
For other researchers, we strongly \blue{recommend} the use of better hardware, as well as a better image-to-text model than we \blue{used}.\footnote{To run BLIP, we only had a MacBook Pro available (with a M1 chip, 8 cores and 8 \sixgigabytes{} of unified memory).}
As a final proposal, maybe the use of multiple image-to-text models could be considered, possibly even including audio-to-text ones.
The outputs of these models could be fed into a large language model, so it can generate a new text that averages and encompasses the semantic meaning all descriptions, improving the chances of ending up with an appropriate caption.

%% file: Publicacion/Conclusiones.tex
\section{Conclusions}\label{sec:Conclusiones}



In this study, we tackled the significant challenge of generating high-quality multimodal datasets, specifically focusing on audio-image-text observations derived from videos. Our motivation stemmed from the increasing demand for diverse and large-scale datasets in the machine learning community, particularly for multimodal data that includes audio, which is often scarce \cite{Audio-to-Image_Cross-Modal, I_Hear_Your_True_Colors}.

We proposed a method to generate these datasets by leveraging continuous video recordings, ensuring a strict semantic connection between \blue{acoustic} and visual data (i.e. both audio and image in each pair are extracted from and related to the same situation). This approach addresses the common issue of undesirable entries in third-party datasets \cite{Large_image_datasets} and the lack of datasets for specific tasks, such as medical image analysis \cite{A_survey_on_Image_Data_Augmentation_for_Deep_Learning}, reinforcement learning \cite{GenRL}, or audio-text in general \cite{I_Hear_Your_True_Colors, BLAT, BLAP, AudioSetCaps}.

Our method involved three key steps: collecting suitable videos, extracting audio-image pairs, and generating textual descriptions for each pair using the BLIP model \cite{BLIP}. This process resulted in the creation of over 2 million audio-image pairs, which were further extended to include textual descriptions, forming a comprehensive multimodal dataset. Despite some limitations, such as the inclusion of frames with text or blurry images, as well as the basic configuration used for text generation, our dataset represents an advancement in the availability of multimodal data for research purposes.

The literature review highlighted the potential of exploiting relationships between audio, image, and text data \blue{\cite{Remote_Sensing_Image_Generation_From_Audio, Deep_Audio-visual_Learning, A_survey_of_multimodal_deep_generative_models, Multimodal_Image_Synthesis_and_Editing}}. These relationships can enhance various applications, including multimodal data analysis, correction of low-quality recordings, video generation, augmented reality, and transfer learning with multimodal models.

Our research underscores the importance of minimizing data modality conversions to preserve data quality. We also emphasized the need for more research in audio-image and audio-text tasks, given the current lack of high-quality data and guidelines for new researchers.

Future work could focus on refining the filtering process to exclude undesirable frames more effectively\blue{, ensuring the temporal alignment by incorporating more recent techniques in the pipeline (such as the ones seen in \cite{AlignNet, Temporal_Alignment_Networks, Temporally_Aligned_Audio_for_Video}),} employing more advanced image-to-text models to improve the quality of textual descriptions, \blue{and even leveraging future audio-to-text models, which could complement the aforementioned descriptions}. Additionally, exploring the potential of incorporating complementary data from any-to-any models, while being mindful of the concerns related to data modality conversions, could further enhance the utility of these datasets.

Overall, our contributions provide a valuable resource for the research community and highlight the importance of multimodal data in advancing machine learning models. We hope that our work will inspire further research and development in this area, ultimately leading to more robust and versatile AI systems.